# Improvement of thermo-mechanical position stability of the beam position monitor in PLS-II

Taekyun Ha, Mansu Hong, Hyuckchae Kwon, Hongsik Han and Chongdo Park

*Pohang Accelerator Laboratory, POSTECH, Pohang, Gyeongbuk 37673, Korea*

In the storage ring of PLS-II, we reduced mechanical displacement of electron beam position monitors (e-BPMs) that is caused by heating during e-beam storage. The orbit feedback system intends that the electron beam pass through the center of the BPM, so to provide stable photon beam into beamlines the BPM pickup itself must be stable to sub-micrometer precision. Thermal deformation of the vacuum chambers on which the BPM pickups are mounted is inevitable when the electron beam current is changed by unintended beam abort. We reduced this deformation by improving the vacuum chamber support and by enhancing the water cooling. We report the thermo-mechanical analysis and displacement measurements of BPM pickups after the improvements.

Email: cdp@postech.ac.kr

Fax: +82-54-279-1799

# I. INTRODUCTION

The Pohang Light Source upgraded machine (PLS-II) started user operation after 6 months of commissioning in 2011 [1, 2]. PLS-II storage ring vacuum system was designed to maintain pressure of $10^{-15}$ bar to achieve beam-gas scattering lifetime $> 20$ h. The vacuum components, especially photon absorbers and rf shielded bellows were designed to endure 3-GeV and 400-mA beam operation [3]. The vacuum system design was also focused on providing a stable photon beam.

To provide a bright photon beam into the beamline, the electron beam size in the storage ring should be as small as possible. The designed electron beam size of PLS-II in a short straight section is 12 μm in the vertical direction, and the orbit must be stable to sub-micrometer precision because beam orbit instability should be $< 5{\sim}10$ % of the beam size. One factor that has a strong influence on this requirement is the mechanical position stability of the e-BPMs that are mounted on vacuum chambers. Because the orbit feedback system is programmed to make the electron beam pass through the center of the BPM, the mechanical stability of the BPM pickup itself is also important [4].

The PLS-II vacuum chamber support near the e-BPM pickup was designed to keep it immovable with respect to a reference plane of the beam orbit. The support made of stainless steel bars is bolted very tightly to the bottom of the vacuum chamber. This strong support was expected to prevent position change of the BPM by external force. During the design stage, thermo-mechanical deformation due to change of electron beam current was not considered significant because PLS-II was intended to operate in top-up mode, in which electrons are frequently injected into the storage ring to make up for decay of the stored beam, and to give a constant electron beam current.

However, the original design of the vacuum chamber and support of PLS-II has two flaws (Fig. 1). First, when temperature increases, the vacuum chamber expands upward because the fixed support is mounted on the bottom of the chamber; as a result, the physical center of the BPM pickup moves. Unintended beam abort during operation can also cause thermo-mechanical deformation of the vacuum chamber because the electron beam current is abruptly changed from 400 mA to 0 mA in $< 1$ s. Second, the temperature of the vacuum chamber takes ~3 h to equilibrate after ring is re-filled, so the position of the BPM pickup stabilizes slowly even if the beam current recovers quickly after a beam abort.

## II. DESIGN OF NEW VACUUM CHAMBER AND SUPPORT

In August of 2015, one sector of the storage ring was changed to a new vacuum system according to the plan of new Elliptically Polarized Undulator (EPU) installation. The photon beam size and flux from this new EPU are much larger than from the existing one, so the vacuum system (including bending magnet chamber and photon absorbers) located downstream of this insertion device were replaced with a new robust one. In addition to compatibility with the new EPU, plans were also made to improve the mechanical stability of this part.

The most important requirement for thermo-mechanical stability is good cooling efficiency for the vacuum chamber. The cooling channel for the existing chamber consists of a copper pipe mounted on the grooves of the vacuum chamber; this indirect cooling scheme has low efficiency. Another problem is the one-sided cooling with respect to the electron beam channel. The unbalanced heat dissipation by this cooling configuration can cause non-uniform deformation of the BPM chamber. We re-designed the vacuum chamber to include cooling channels on both sides of the beam channel for direct and uniform cooling; we also reformed the support for the BPM chamber to expand equally in the vertical direction when heated.

The new vacuum chamber and support system undergo much less thermo-mechanical expansion with than the existing ones (Fig. 2). The center of the BPM in the existing vacuum chamber rises when heated (Fig. 2a) because the chamber can only expand upward due to the fixed support that holds up the chamber from the bottom. In contrast the center of the BPM pickup in the new design stays in the original position regardless of thermal load (Fig. 2b).

## III. IMPROVEMENT OF MECHANICAL STABILITY OF THE BPM PICKUP

We calculated thermo-mechanical effect of both types of vacuum chamber and support, then compared thermal displacement, and time to thermal equilibrium. The storage ring vacuum chamber is heated mainly by synchrotron radiation that is reflected from photon absorbers, and by electron beam image current. The synchrotron radiation power from a bending magnet of the PLS-II storage ring is ~ 17.5 kW when 400-mA beam is stored; about 20% is reflected onto the vacuum chamber. In a Multipole-II chamber of PLS-II, taking into account that the length ratio of this chamber is only 1.3 m/282 m, we assumed a uniform heat load of 26 W (16 W from the synchrotron radiation + 4 W from the image current + 6 W from HOM heating of BPM pickups) being dissipated onto the inner surface of the electron beam chamber. The boundary condition for cooling hole of the new chamber is set to 0.01 W/(mm$^2$·K), which is the typical convection coefficient of water, but for the existing chamber where the copper pipe is mounted on the outside groove of the chamber the convection coefficient was assumed to be 10 % of that in the case with perfect contact. This assumption was based on the observation that the contact area between the cooling pipe and the chamber groove is ≤10 % of the groove area.

Thermo-mechanical analysis indicated that displacements of the BPM pickup were smaller for the new designs than for the existing ones (Fig. 3). The BPM pickup in the old vacuum chamber moved about 34.5 μm vertically when heated (Fig. 3a), but that in the new vacuum chamber moved only about 6.5 μm (Fig. 3b). A linear variable differential transducer (LVDT) was used to measure the actual position of the top of the BPM. When the electron beam was abruptly decreased from 300 mA to 0 mA, the top of the existing chamber dropped 25 μm, but that of the new chamber dropped only 3.5 μm (Fig.4). The difference between the simulated and measured values is due to the difference in the stored current (400 mA or 300 mA) but the new chamber and support design is effective in both results.

This displacement can also be reduced by adjusting the location of the cooling channel and the contact between the chamber and the support. In the existing system, the contact points of the chamber and the support are just below the beam channel where heating is not negligible (Fig. 2), so a small amount of the heat is transmitted to the support and therefore the support itself also expands upward direction; this expansion amplifies the displacement. However, in the new chamber, the contacts between the chamber and the support are outside of the cooling channel. In this arrangement, very little heat is transferred from the chamber to the support.

The increased cooling efficiency new system reaches thermal equilibrium after sudden change in thermal load much faster than the old one (Fig. 5). When a beam was aborted suddenly due to a machine interlock from some reason and the beam current of the storage ring was recovered within 30 min, the new vacuum chamber took 1 h to reach thermal equilibrium whereas the old one took 3 h. During this, the photon beam provided into the beamline also drifted for 3 h. This long transition time is due to the insufficient cooling efficiency of the system.

## IV. SUMMARY

We improved mechanical stability of the BPM in the PLS-II storage ring by modifying the vacuum chamber and the support. The new vacuum chamber was made of extruded aluminum chamber with cooling channels on both sides of the electron beam channel. The cooling efficiency of the new chamber was excellent, so that it reduced the time to reach thermal equilibrium to 1 h from 3 h. The thermal expansion in the new design is allowed in both upward and downward directions, so the new support also decreases BPM displacement during heating. Results of thermo-mechanical analysis agree well with measured values. This design will be used in all storage ring vacuum chambers and supports.

## ACKNOWLEDGEMENT

This research was supported by the Converging Research Center Program through the Ministry of

Figure Captions.

Figure 1. Typical behavior of BPM pickup displacement on the bending magnet chamber after an abrupt beam abort. Beam current is recovered within half an hour, but the temperature or displacement of the BPM stabilizes after 3 h, so the photon beam is unstable.

Figure 2. Schematics of the thermo-mechanical deformation of (a) the existing support design, and (b) the new support design.

Figure 3. Thermo-mechanical analyses of the BPM pickup for (a) the existing chamber and support, and (b) the new chamber and support for a multipole-II chamber of PLS-II.

Figure 4. Measured BPM pickup displacement after an abrupt beam abort from 300 mA.

Figure 5. Required time to reach equilibrium after an abrupt beam abort followed by quick recovery of the beam current.

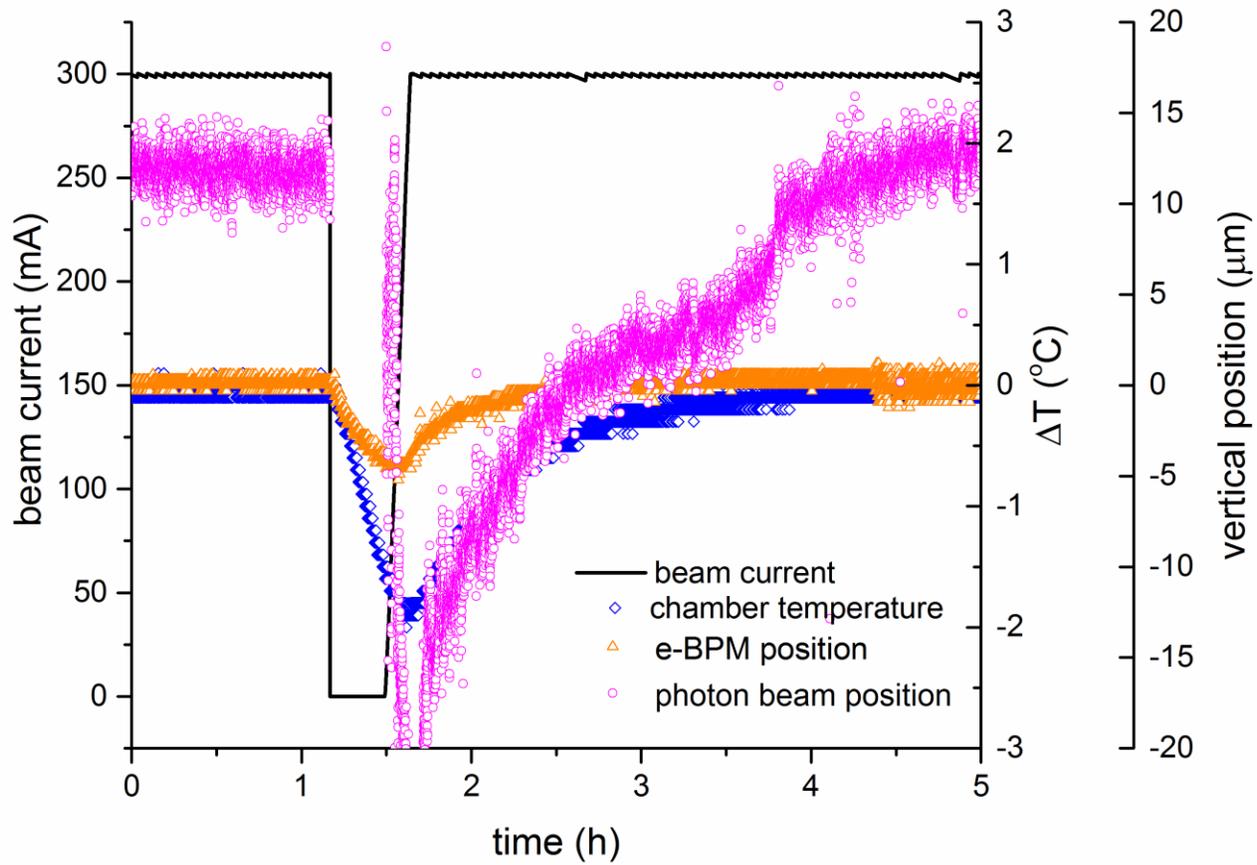

Figure 1

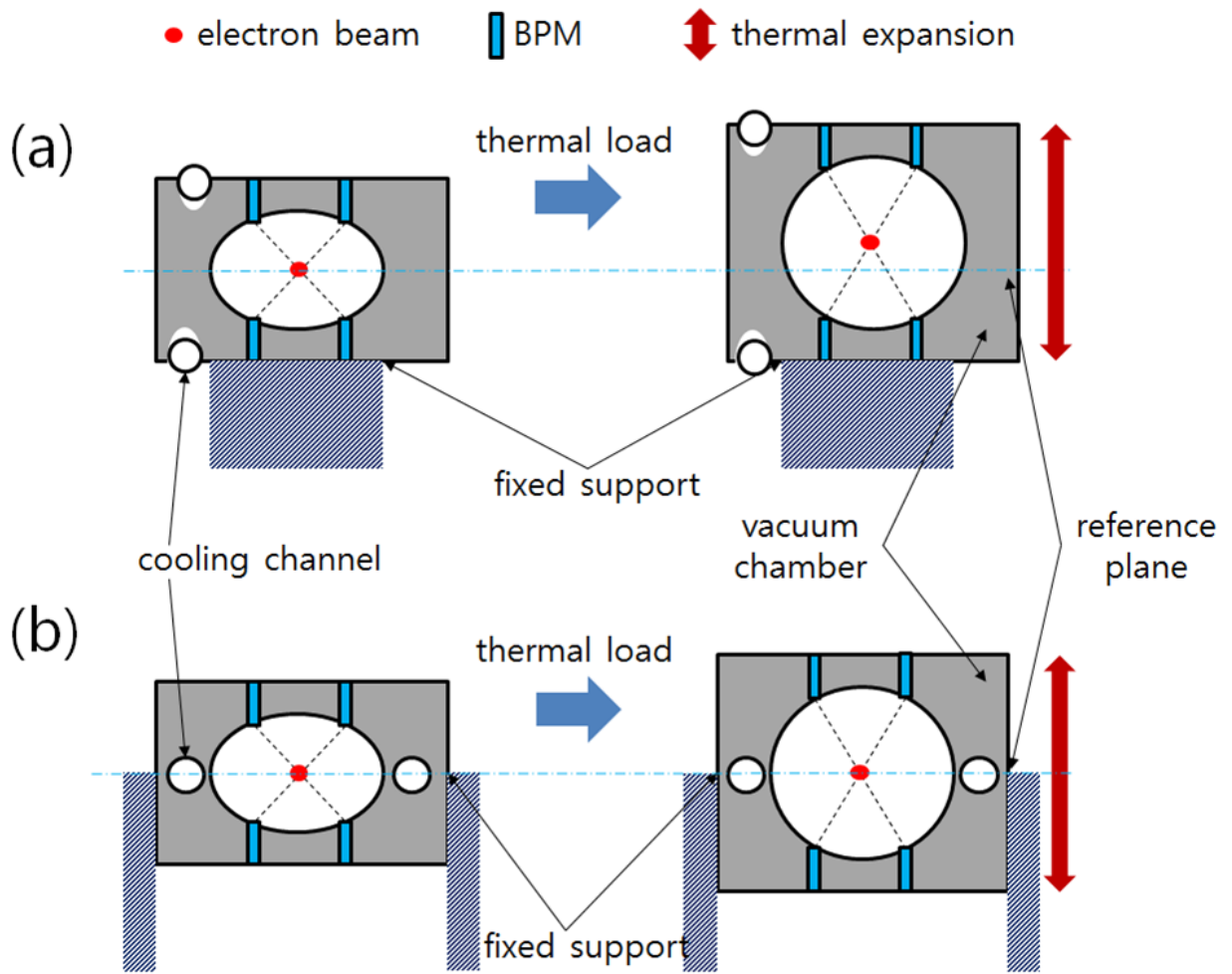

Figure 2

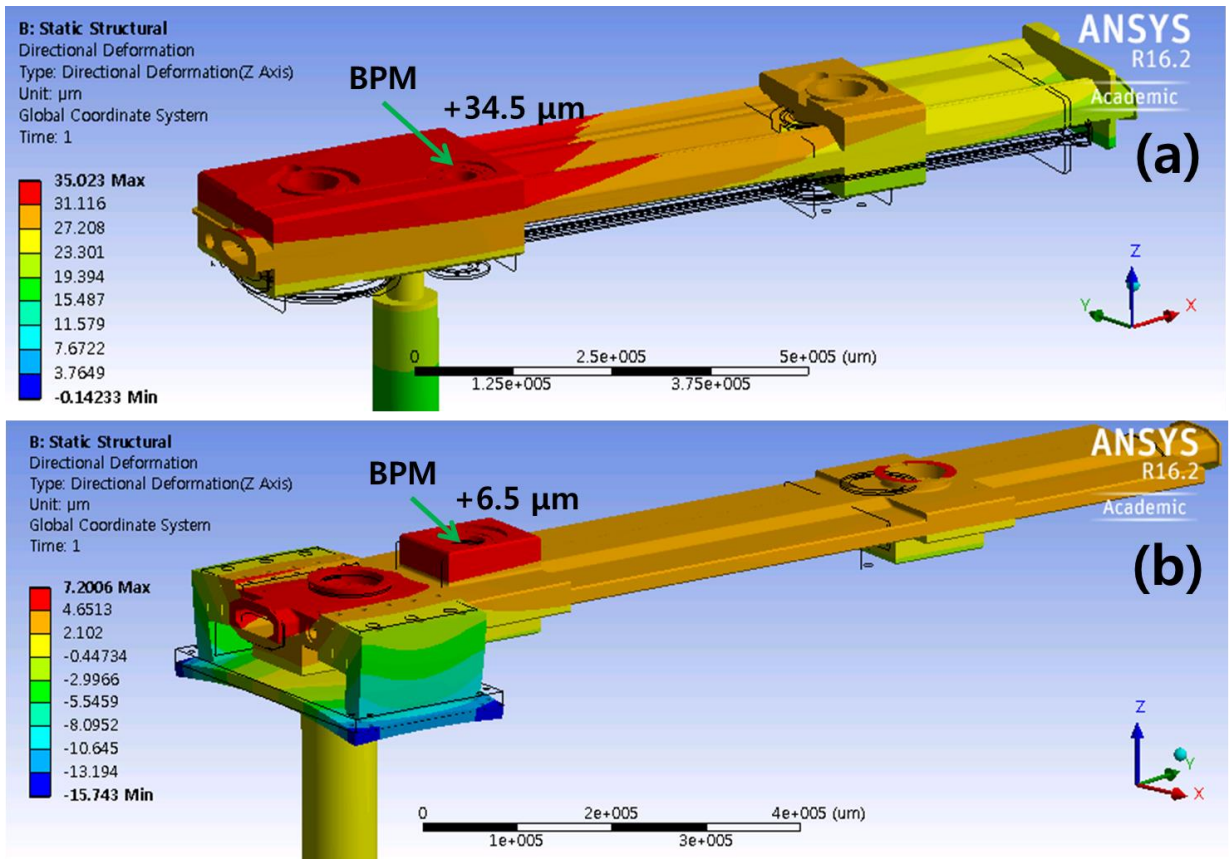

Figure 3

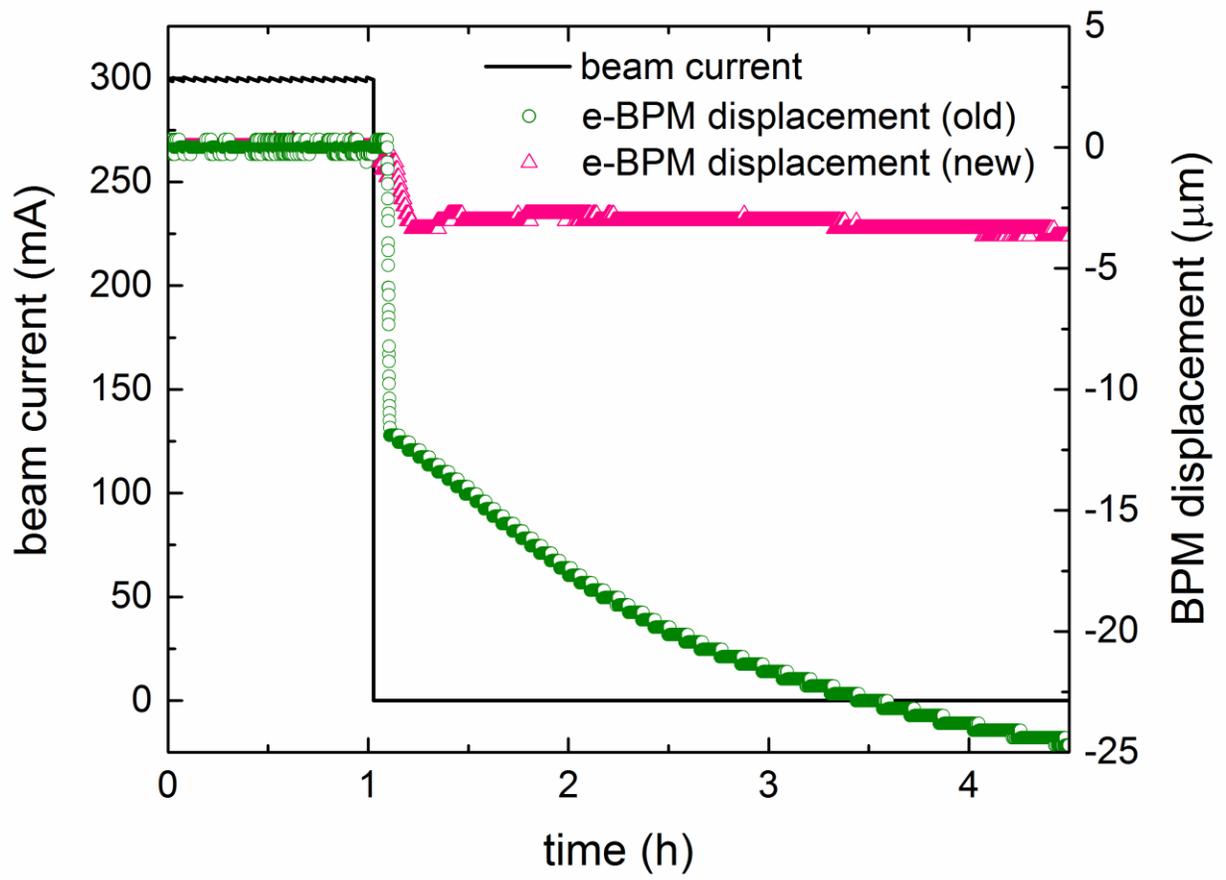

Figure 4

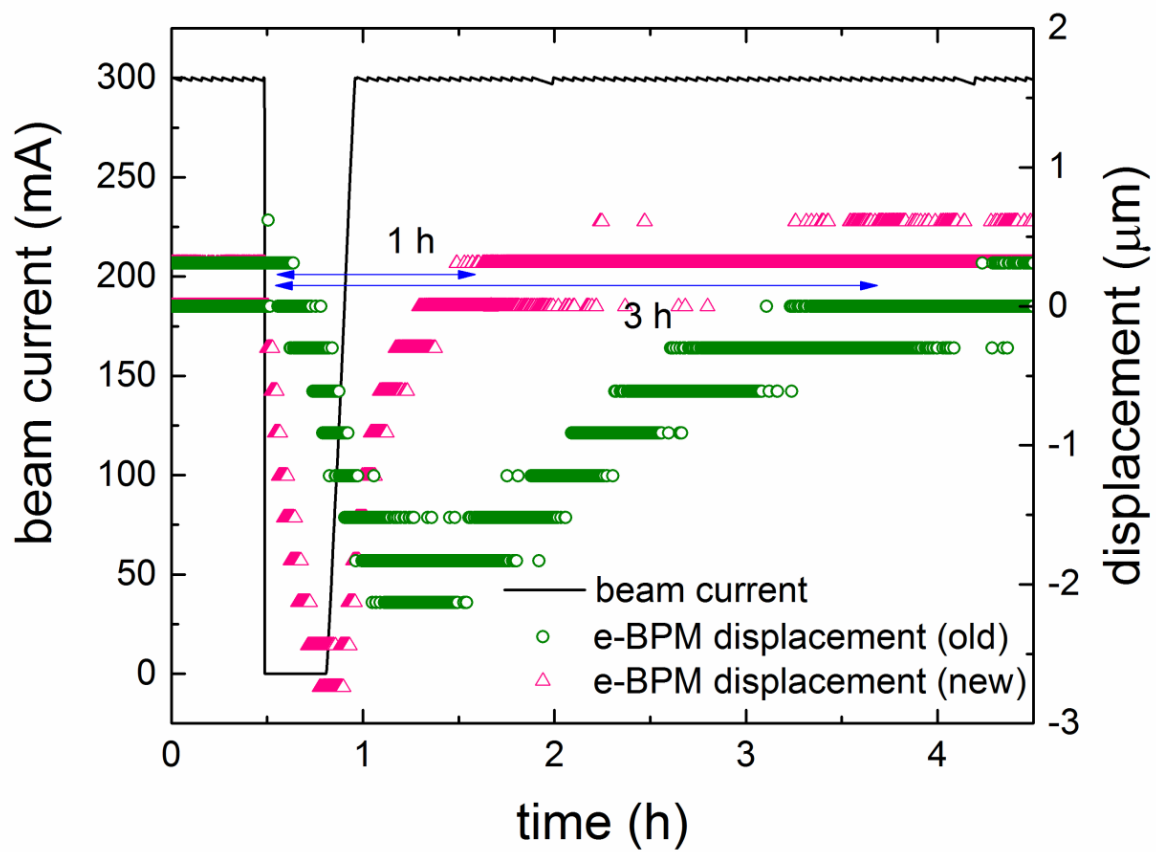

Figure 5